\documentstyle[preprint,cite,aps]{revtex}
\newcommand{\be}{\begin{equation}}
\newcommand{\ee}{\end{equation}}
\newcommand{\bea}{\begin{eqnarray}}
\newcommand{\eea}{\end{eqnarray}}
\newcommand{\ba}{\begin{array}}
\newcommand{\ea}{\end{array}}

\newcommand{\bb}{\bibitem}

\begin{document}
\draft

\title{\bf New renormalization group approach and scaling laws for the Lifshitz critical behavior}
\author{Marcelo M. Leite\footnote{e-mail:leite@fis.ita.br}}
\address{{\it Departamento de F\'\i sica, Instituto
Tecnol\'ogico de Aeron\'autica, Centro T\'ecnico Aeroespacial, 
12228-900, S\~ao Jos\'e dos Campos, SP, Brazil}}
\maketitle

\vspace{0.5cm}

\begin{abstract}
{\it A new renormalization group treatment is proposed for the critical 
exponents of an $m$-fold Lifshitz point. The anisotropic cases ($m\neq8$) 
are described by two independent fixed points associated to two independent 
momentum flow along the quadratic and quartic directions, respectively. The 
isotropic case is described separately. In that case, the fixed point is due 
to renormalization group transformations along the quartic directions. The new 
scaling laws are derived for both cases and generalize the ones previously reported.}
\end{abstract}

\vspace{3cm}

\newpage

Since the early formulation of the $m$-axial Lifshitz critical behavior 
associated to the onset of helical order 
in magnetic systems \cite{1,2}, there have been various studies pointing out 
other applications of this kind of critical behavior in other real physical 
systems like high-$T_c$ superconductors \cite{3}, 
ferroelectric liquid crystals \cite{4}, magnetic materials and alloys \cite{Bece} etc. 
These multicritical points 
appear at the confluence of a disordered
phase, a uniformly ordered phase and a modulated
ordered phase \cite{1,2}. The modulated phase is 
characterized by a fixed equilibrium wavevector which 
goes continuously to zero as the system 
approaches the Lifshitz point. In case this
wavevector has $m$-components, the critical system presents an $m$-fold 
Lifshitz critical behavior. If $m=d$, it displays an isotropic critical 
behavior. There are 
several  types of isotropic behavior, but we will be concerned here 
only with the case  $m=d$ near to 8. Otherwise ($m\neq d\neq8$), 
the system presents an anisotropic critical behavior. The particular interest in the 
isotropic case $m=d$ near to 8 is due to its similarity to the anisotropic case, for 
both ones can be studied using the same field-theoretic description.

The purpose of this letter is to present new renormalization group (RG) arguments 
and new scaling laws that point towards a natural explanation of the $m$-fold 
Lifshitz critical behavior for the anisotropic and isotropic cases. Our 
reasoning will be based on magnetic systems. The $m$-fold Lifshitz point can 
be described by a generalization of the ANNNI model \cite{Selke}, 
which is a spin-$\frac{1}{2}$ Ising model on a $d$-dimensional lattice 
with nearest-neighbors interacting ferromagnetically as well next-nearest-neighbors with antiferromagnetic 
couplings along $m$ directions. Its field theoretical representation is given in terms 
of a modified $\phi^4$ theory with higher derivative terms along the $m$ directions, with 
the following bare Lagrangian density \cite{AL1, 8}:

\begin{equation}
L = \frac{1}{2}|\bigtriangledown_{m}^{2} \phi_0\,|^{2} + 
\frac{1}{2}|\bigtriangledown_{(d-m)} \phi_0\,|^{2} + 
\delta_0  \frac{1}{2}|\bigtriangledown_{m} \phi_0\,|^{2} 
+ \frac{1}{2} t_{0}\phi_0^{2} + \frac{1}{4!}\lambda_0\phi_0^{4} .
\end{equation}

At the Lifshitz point $\delta_0 = t_0 = 0$. The renormalized correlation 
functions are defined at the critical theory for $t=0$, $M=0$, where 
$t = Z_{\phi^2}^{-1} t_0$, $M = Z_{\phi}^{\frac{-1}{2}} \phi_0$ are the renormalized reduced 
temperature and order parameter (magnetization), respectively . Below the 
Lifshitz critical temperature $T_{L}$, one can expand the renormalized vertex 
parts for $t \neq 0$, $M \neq 0$ around the ones for $t=0$, $M=0$ as a power series 
in $t$ and $M$, whereas above  $T_{L}$ the renormalized vertices for $t \neq 0$ 
are expanded around the renormalized parts calculated at $t=0$ in powers of $t$  \cite{Amit}.   
We shall keep $\delta_0 = 0$ from now on, since this choice for the renormalized theories 
simplifies the dimensional analysis. The dependence on the external momenta  along the 
quadratic (noncompeting) and quartic (competing) directions can be split in two independent 
contributions. The RG flow in the quartic momenta scale can be done separately, following a 
situation described by Wilson in the early seventies \cite{wilson}. 
The absence of quadratic terms in the 
components of the momenta along the competition axes allows us to perform a 
dimensional redefinition of the $m$-dimensional momenta subspace. This is the main advantage of 
choosing $\delta_{0}=0$ as the starting point for our RG analysis.

Our new point is to consider the momenta along the competing directions 
entering with half the power of a momentum scale in dimensional analysis. 
For the anisotropic case, the momenta 
scales associated to the two orthogonal subspaces, namely the ($d-m$)-dimensional 
noncompeting one, characterized by the renormalization scale $\kappa_1$, and the competing $m$ 
directions chacterized by $\kappa_2$, flow independently under RG transformations. 
This produces two independent fixed points, which lead to two new independent sets of 
scaling laws for each subspace. These new relations are valid to arbitrary order in 
an $\epsilon_L$-expansion, where $\epsilon_L = 4 + \frac{m}{2} - d$. In particular, the conventional 
Josephson hyperscaling relation for the anisotropic $m$-fold Lifshitz critical behavior \cite{1},

\begin{equation}
2 - \alpha_{L} = (d-m)\nu_{L2} + m \nu_{L4} ,
\end{equation}
should be replaced by two independent new relations, namely the one associated to 
correlations perpendicular to the competition axes  

\begin{equation}
2 - \alpha_{L2} = (d-\frac{m}{2})\nu_{L2} ,
\end{equation}
together with the hyperscaling law associated to correlations along the competing 
directions

\begin{equation}
2 - \alpha_{L4} = 2 (d-\frac{m}{2})\nu_{L4} .
\end{equation}

Furthermore, the other new scaling relations appropriate for the  critical exponents 
perpendicular to the competition axes,
\begin{mathletters}
\begin{eqnarray}
\gamma_{L2} &=& \nu_{L2} (2 - \eta_{L2}), \label{1}\\
\delta_{L2} &=& \frac{(d-\frac{m}{2}) + 2 - \eta_{L2}}{(d-\frac{m}{2}) - 2 + \eta_{L2}}, \label{2}\\
\beta_{L2} &=& \frac{1}{2} \nu_{L2} ((d-\frac{m}{2}) - 2 + \eta_{L2}), \label{3}
\end{eqnarray}
\end{mathletters}
imply the Widom $\gamma_{L2} = \beta_{L2} (\delta_{L2} -1)$ and 
Rushbrook $\alpha_{L2} + 2 \beta_{L2} + \gamma_{L2} = 2$ relations. The ones associated to 
critical correlations along the competition axes are analogous, namely
\begin{mathletters}
\begin{eqnarray}
\gamma_{L4} &=& \nu_{L4} (4 - \eta_{L4}), \label{4}\\
\delta_{L4} &=& \frac{2 (d-\frac{m}{2}) + 4 - \eta_{L4}}{2 (d-\frac{m}{2}) - 4 + \eta_{L4}}, \label{5}\\
\beta_{L4} &=& \frac{1}{2} \nu_{L4} (2 (d-\frac{m}{2}) - 4 + \eta_{L4}), \label{6} 
\end{eqnarray}
\end{mathletters}
which lead to $\gamma_{L4} = \beta_{L4} (\delta_{L4} -1)$ and 
$\alpha_{L4} + 2 \beta_{L4} + \gamma_{L4} = 2$. 
Note that at one-loop level, $\nu_{L4} = \frac{1}{2} \nu_{L2}$, and 
the above equations reduce to the scaling relations given in Ref. \cite{1}.  
In addition, for the isotropic behavior $m=d$ 
near to $8$ (the expansion parameter is $\epsilon_{L} = 8-d$) and the 
new corresponding scaling relations are:
\begin{mathletters}
\begin{eqnarray}
2 - \alpha_{L4} &=& d \nu_{L4} ,\label{7}\\
\gamma_{L4} &=& \nu_{L4} (4 - \eta_{L4}) , \label{8}\\
\delta_{L4} &=& \frac{d + 4 - \eta_{L4}}{d - 4 + \eta_{L4}}, \label{9}\\
\beta_{L4} &=& \frac{1}{2} \nu_{L4} (d - 4 + \eta_{L4}), \label{10}
\end{eqnarray}
\label{11}
\end{mathletters} 
which give rise to the same Widom and Rushbrook scaling relations mentioned 
above.

Let us describe the method in some detail. We start with the anisotropic case. 
We choose two independent sets of normalization conditions for the critical 
theory, i.e., two different (subtraction) symmetry points. The first one is 
chosen with nonvanishing external momenta along the noncompeting 
$(d-m)$-dimensional subspace. The theory is renormalized at (quadratic) external  
momenta scale $\kappa_{1}$. The flow in this scale gives origin to 
the critical indices $\nu_{L2}$ and  $\eta_{L2}$.
The second one is defined at nonvanishing external 
momenta along the $m$-dimensional competing (quartic) subspace. The theory is 
renormalized at (quartic) external  momenta scale $\kappa_{2}$, 
originating the critical exponents  $\nu_{L4}$ and  $\eta_{L4}$. Therefore, 
there are two sets of renormalized vertex parts, characterized by the scales 
$\kappa_{1}$ and $\kappa_{2}$, defined by:

\begin{equation}
\Gamma_{R(\tau)}^{(N,L)} (k_{i(\tau)}, p_{i(\tau)}, g_{\tau}, \kappa_{\tau}) = 
Z_{\phi (\tau)}^{\frac{N}{2}} Z_{\phi^{2} (\tau)}^{L} 
\Gamma_{(\tau)}^{(N,L)} (k_{i(\tau)}, p_{i(\tau)}, \lambda_{\tau}, \Lambda_{\tau}) ,
\end{equation}
where $g_{\tau}$ are the renormalized coupling constants, $Z_{\phi (\tau)}$ 
and $Z_{\phi^{2} (\tau)}$  are the field and temperature 
normalization constants and $\Lambda_{\tau}$ are the associated momentum cutoffs along the 
quadratic and quartic directions, respectively. (Except for $\Gamma_{R(\tau)}^{(0,2)}$ which 
cannot be renormalized multiplicatively, all other vertex parts can be managed within 
this argument.) The label  $\tau = 1 (2)$ 
refers to the nonvanishing quadratic (quartic) external momenta and zero quartic (quadratic) 
external momenta. As usual, $k_{i}$ ($i=1,...,N$) are the momenta associated to the vertex functions 
$\Gamma_{R}^{(N,L)}$ with $N$ external legs and $p_{i}$ ($i=1,...,L$) are the momenta associated to the 
$L$ insertions of $\phi^{2}$ operators. 
In terms of dimensionless parameters, $g_{\tau} = u_{\tau} (\kappa_{\tau}^{2 \tau})^{\frac{\epsilon_{L}}{2}}$, 
and $ \lambda_{\tau} =  u_{0 \tau} (\kappa_{\tau}^{2 \tau})^{\frac{\epsilon_{L}}{2}}$. As 
usual, $u_{0 \tau}$, $Z_{\phi (\tau)}$ and $Z_{\phi^{2} (\tau)}$ can be represented as 0
power series in $u_{\tau}$. (Henceafter, we shall suppress the cutoff $\Lambda_{\tau}$.) 
Invariance of the bare functions under the momenta scales used to fix the renormalized 
theories leads to the RG equation for $\Gamma_{R(\tau)}^{(N,L)}$

\begin{equation}
(\kappa_{\tau} \frac{\partial}{\partial \kappa_{\tau}} + 
\beta_{\tau}\frac{\partial}{\partial u_{\tau}} 
- \frac{1}{2} N \gamma_{\phi (\tau)}(u_{\tau}) + L \gamma_{\phi^2 (\tau)}(u_{\tau})) 
\Gamma_{R (\tau)}^{(N,L)} = \delta_{N,0} \delta_{L,2} (\kappa_{\tau}^{-2 \tau})^\frac{\epsilon_{L}}{2} 
B_{\tau}(u_{\tau}) ,      
\end{equation}
where $\beta_{\tau} = (\kappa_{\tau}\frac{\partial u_{\tau}}{\partial \kappa_{\tau}})$, 
$\gamma_{\phi (\tau)}(u_{\tau})  = \beta_{\tau} 
\frac{\partial ln Z_{\phi (\tau)}}{\partial u_{\tau}}$ 
and $\gamma_{\phi^{2} (\tau)}(u_{\tau}) = - \beta_{\tau} 
\frac{\partial ln Z_{\phi^{2} (\tau)}}{\partial u_{\tau}}$ are calculated at fixed 
bare coupling $\lambda_{\tau}$. In terms of the bare (dimensionless) coupling constant, 
$\beta_{\tau} = - \tau \epsilon_{L} (\frac{\partial ln u_{0 \tau}}{\partial u_{\tau}})^{-1}$. 
The volume element in momenta space associated to loop  
integrals, $d^{d-m}q d^{m}k$, has engineering dimension  
$\Lambda^{d - m} \Lambda^{\frac{m}{2}} = \Lambda^{d - \frac{m}{2}}$, where 
$\vec{q} ([\vec{q}] = \Lambda)$   represents a 
$(d-m)$-dimensional vector perpendicular to the competing directions 
and $\vec{k} ([\vec{k}] = \Lambda^{\frac{1}{2}})$
an $m$-dimensional vector along the competing axes, respectively. Under a flow in the external momenta 
$k_{i (\tau)}$, we have the following simple scaling properties at the fixed points $u_{\tau}^{*}$ 
$((N,L) \neq (0,2))$ :
\begin{equation}
\Gamma_{R (\tau)}^{(N,L)} (\rho k_{i (\tau)}, \rho p_{i (\tau)}, u_{\tau}^{*}, \kappa_{\tau}) = 
\rho^{\tau [N + (d - \frac{m}{2}) - \frac{N(d - \frac{m}{2})}{2} -2L] - 
\frac{N \gamma_{\phi (\tau)}^{*}}{2} + L \gamma_{\phi^{2} (\tau)}^{*}} 
\Gamma_{R (\tau)}^{(N,L)} (k_{i (\tau)}, p_{i (\tau)}, u_{\tau}^{*}, \kappa_{\tau}). 
\end{equation}

Now take $L=0$. Above the Lifshitz critical temperature, the RG equation takes the form:
\begin{equation}
(\kappa_{\tau} \frac{\partial}{\partial \kappa_{\tau}} + 
\beta_{\tau}\frac{\partial}{\partial u_{\tau}} 
- \frac{1}{2} N \gamma_{\phi (\tau)}(u_{\tau}) + 
\gamma_{\phi^2 (\tau)}(u_{\tau}) t\frac {\partial}{\partial t}) 
\Gamma_{R (\tau)}^{(N)} = 0.
\end{equation}
At the fixed point $\beta_{\tau}(u_{\tau}^{*})=0$, and the solution can be written as:
\begin{equation}
\Gamma_{R (\tau)}^{(N)} (k_{i (\tau)}, t, u_{\tau}^{*}, \kappa_{\tau})=
\kappa_{\tau}^{\frac{N \gamma_{\phi (\tau)}^{*}}{2}} 
F_{(\tau)}^{(N)}(k_{i (\tau)},\kappa_{\tau} t^{\frac{-1}{\gamma_{\phi^{2} (\tau)}^{*}}}) .
\end{equation}
Define $\eta_{\tau} = \gamma_{\phi (\tau)}^{*}$, $\theta_{\tau} = -\gamma_{\phi^{2} (\tau)}^{*}$, 
such that $\eta_{1} = \eta_{L2}$, $\eta_{2} = \eta_{L4}$, and so on. Under a flow in the external momenta, 
we have:

\begin{eqnarray}
\Gamma_{R (\tau)}^{(N)} (k_{i (\tau)}, t, \kappa_{\tau}) =&&
\rho^{\tau[N + (d- \frac{m}{2}) - \frac{N}{2}(d - \frac{m}{2})] -\frac{N}{2} \eta_{\tau}} 
\kappa_{\tau}^{\frac{N}{2} \eta_{\tau}} \nonumber\\
&&F_{(\tau)}^{(N)}(\rho^{-1} k_{i (\tau)},(\rho^{-1}\kappa_{\tau}) 
(\rho^{-2 \tau}t)^{\frac{-1}{\theta_{\tau}}} ) . 
\end{eqnarray}

By choosing $\rho = \kappa_{\tau} (\frac{t}{\kappa^{2 \tau}})^{\frac{1}{\theta_{\tau} + 2 \tau}}$, and 
replacing back in (13), the vertex function depends only on the combination $k_{i} \xi$ apart from a 
power of $t$. Therefore, we can identify the critical exponents $\nu_{1} = \nu_{L2}$ and  $\nu_{2} = \nu_{L4}$ as 
\begin{equation}
\nu_{\tau}^{-1} = 2 \tau + \theta_{\tau} .
\end{equation}

Hence, at the fixed point all correlation functions (not including composite operators) scale at $T>T_{L}$, 
since they are functions of  $k_{i} \xi$ only. For $N=2$ we choose $\rho = k_{i (\tau)}$, the external momenta. 
As  $k_{i (\tau)} \rightarrow 0$, 
$\Gamma_{R}^{(2)} = \chi^{-1}$ and we can identify the susceptibility critical exponent:
\begin{equation}
\gamma_{\tau} = \nu_{\tau} (2 \tau - \eta_{\tau}),
\end{equation}
which are Eqs. (\ref{1}) and (\ref{4}). Below $T_{L}$, the renormalized equation of state at the fixed point 
is expanded in powers of $M$, which under a flow in the momenta turns out to be
\begin{equation}
H(t, M, \kappa_{\tau}) = \rho^{\tau [\frac{d - \frac{m}{2}}{2} + 1]}\nonumber\\
\;\; H(\frac{t}{\rho^{2 \tau}}, \frac{M}{\rho^{2 \tau [\frac{d - \frac{m}{2}}{2} - 1]}}, \frac{\kappa_{\tau}}{\rho}) .
\end{equation}
We choose $\rho$ to be a power of $M$:
\begin{equation}
\rho = \kappa_{\tau} [\frac{M}{\kappa_{\tau}^{\frac{\tau}{2}[(d - \frac{m}{2}) - 2]}}]^{\frac{2}
{\tau[(d - \frac{m}{2}) - 2] + \eta_{\tau}}} ,
\end{equation}
and from the scaling form of the equation of state 
$H(t, M) = M^{\delta_{\tau}} f(\frac{t}{M^{\frac{1}{\beta_{\tau}}}})$, Eqs. (\ref{2}), (\ref{3}), (\ref{5}), (\ref{6}) 
follow in a straightforward manner.

The specific heat exponents can be obtained by analysing the RG equation for 
$\Gamma_{R (\tau)}^{(0,2)}$ above $T_{L}$ at the fixed point, i.e.
\begin{equation}
(\kappa_{\tau} \frac{\partial}{\partial \kappa_{\tau}} + 
\gamma_{\phi^2 (\tau)}^{*} (2 + t\frac {\partial}{\partial t})) 
\Gamma_{R (\tau)}^{(0,2)} =  (\kappa_{\tau}^{-2 \tau})^{\frac{\epsilon_{L}}{2}} 
B_{\tau}(u_{\tau}^{*}) ,    
\end{equation} 
whose solution has the form:
\begin{equation}
\Gamma_{R (\tau)}^{(0,2)}) = 
(\kappa_{\tau}^{-2 \tau})^{\frac{\epsilon_{L}}{2}} 
(C_{1 \tau} (\frac{t}{\kappa_{\tau}^{2 \tau}})^{- \alpha_{\tau}} + 
\frac{\nu_{\tau}}{\tau \nu_{\tau}(d - \frac{m}{2}) -2} B_{\tau}(u_{\tau}^{*})).
\end{equation}
Hence, we can identify $\alpha_{L2} (\tau =1)$ and $\alpha_{L4} (\tau =2)$ 
as equations (3) and (4). 

For the isotropic case, one has to consider only 
the momenta scale $\kappa_{2}$, corresponding to the quartic direction. The 
volume element in momenta space is $\Lambda^{\frac{d}{2}}$. The $\beta$ function, 
given by $\beta(u_{2}) = -\epsilon_{L} \frac{\partial ln u_{02}}{\partial u_{2}}$, 
is different from the anisotropic one along the quartic direction. 
This implies that the isotropic critical behavior cannot be obtained from the 
anisotropic one whatsoever, for the coupling constants in both cases have 
different canonical dimensions. In the sequel, for sake of simplicity, one can replace $(d - \frac{m}{2})$ 
by $\frac{d}{2}$ in the formulae for the critical exponents $\nu_{L4}$ and $\eta_{L4}$ for the anisotropic behavior. 
Proceeding along the same lines, we obtain the scaling relations (\ref{11}). 

From this analysis, it is easy to reproduce the critical exponents at one-loop level given in Ref. \cite{1} along the 
competing axes for both anisotropic and isotropic cases. Note that even though the fixed points are the same 
at one-loop level along the quartic and quadratic flow in the momenta scales for both cases, they do not have to be 
the same at higher loops. It follows naturally that a thorough description of 
the $m$-fold Lifshitz critical behavior actually needs {\it four} independent critical indices for the anisotropic case 
(in order to obtain all of them via scaling relations), reducing to two for the isotropic case. 
Their independence means that if we are willing to make suitable approximations to solve higher loop 
integrals, then they can be done independently in either subspace for the anisotropic case. 

We can check the 
validity of the hyperscaling relation (3) utilizing this reasoning. In the second paper of reference \cite{AL1} 
the exponent $\nu_{L2}$ and $\gamma_{L2}$ were calculated within a new two-loop approximation. For $d=3, m=N=1$ 
(ANNNI model) the $\epsilon_{L}$-expansion yielded $\nu_{L2} = 0.73$ and $\eta_{L2} = 0.04$. The exponent  
$\gamma_{L2}=1.45$ was obtained analytically from the $\epsilon_{L}$-expansion together with the scaling relation.
However, since $\epsilon_{L} = 1.5$ is not a small parameter the neglected $O(\epsilon_{L}^3)$ contribution 
could be relevant to the calculation of $\gamma_{L2}$. In fact, had we replaced the numerical values of 
$\nu_{L2} = 0.73$ and $\eta_{L2} = 0.04$ directly into the scaling law, we would have obtained $\gamma_{L2} =1.43$. 
Therefore, whenever the expansion parameter  $\epsilon_{L}>1$, one should use the {\it numerical} values of the exponents 
$\nu_{L2}$ and $\eta_{L2}$ obtained from the $\epsilon_{L}$-expansion for fixed values of $(m,N,d)$ in order to obtain 
the numerical values of the other exponents via scaling laws. The same reasoning applies when dealing with exponents 
associated to the competing axis, namely $\nu_{L4}$, $\eta_{L4}$, etc. We can test this procedure by calculating other 
critical exponents perpendicular to the competing axis. By replacing the value of $\nu_{L2}$ in (3), we find 
$\alpha_{L2} = 0.175$, which is in remarkable agreement to the 
most recent Monte Carlo output $\alpha_{L2} = 0.18 \pm 0.02$ \cite{henkel}. The susceptibility exponent calculated 
using this numerical method yielded $\gamma_{L2}=1.36 \pm 0.03$. On the other hand, from (\ref{3}), 
$\beta_{L2} = 0.198$, whereas that simulation resulted in $\beta_{L2} = 0.238 \pm 0.005$, which is not 
as good as the result for the specific heat exponent, but is still reasonable. 

Indeed in \cite{henkel}, the authors 
found the exponents along directions perpendicular to the competing axes and identified these exponents with 
$\alpha_{L}$, $\gamma_{L}$ and $\beta_{L}$. After the work presented in this letter, we found appropriate to identify 
them with $\alpha_{L2}$, $\gamma_{L2}$ and $\beta_{L2}$, respectively. 
Nevertheless, a curious fact takes place in the anisotropic case if in any 
given order in perturbation theory $\nu_{L2} = 2 \nu_{L4}$ and 
$\eta_{L2}=\frac{1}{2} \eta_{L4}$, namely 
$\alpha_{L2}= \alpha_{L4} = \alpha_{L}$, 
$\gamma_{L2}= \gamma_{L4} = \gamma_{L}$ and 
$\beta_{L2}= \beta_{L4} = \beta_{L}$. The scaling relations obtained in 
Ref. \cite{1} for the anisotropic case turn out to reduce to those given here 
\cite{Lei}. This shows that the new results displayed here 
are consistent with precise numerical data.  

Previous RG treatments \cite{Me, Di}, introduced a dimensionful constant $\sigma$ in front of the first term in the 
Lagrangian presented in this work. In Ref. \cite{Me} another set of normalization conditions along with two 
symmetry points were chosen in order to renormalize the 1PI vertex parts. Those authors chose the symmetry point at 
nonvanishing quartic external momenta and zero quadratic external momenta for the vertex parts $\Gamma_{R}^{(4)}$, 
$\Gamma_{R}^{(2,1)}$ and $\Gamma_{R}^{(0,2)}$. In addition, they used two conditions on the derivative 
of the two-point vertex function at two independent external momenta scales. They argued that this choice of 
renormalization points would make the bare parameters and renormalization constants $\sigma$-independent. 
This reasoning implicitly takes into account the existence of only one fixed point, since it mixes two different 
symmetry points at the same set of normalization conditions for $\Gamma_{R}^{(2)}$, in conformity with the 
first approach based on a one-loop analysis \cite{1}. These two simultaneous conditions are responsible for the 
identifications $\gamma_{L2} = \gamma_{L4}$, etc. The limitation of such renormalization group 
analysis is evident, since it fails to obtain the isotropic scaling relations. On the other hand, our 
analysis separates the two symmetry points in two independent sets of normalization conditions with independent 
renormalization group invariance in both sectors. Thanks to the condition $\delta_{0}=0$ fulfilled at the Lifshitz 
point, the first term of the above Lagrangian (quartic in the momenta) does not need to be multiplied by a dimensionful 
constant in order to make sense on dimensional grounds. Hence, $\sigma$ is not required at all in our new 
renormalization group approach, which permits freedom and simplicity to tackle the Lifshitz behavior in its full 
generality. 

The scaling relations presented for the isotropic critical behavior resembles the ones for the usual 
Ising-like system, the difference being the critical dimension in both cases. 
Except for the hyperscaling relation (which looks the same as the Ising-like one), all other 
scaling laws can be obtained from the usual $\phi^{4}$ theory by making the replacement $2 \to 4$. 
Besides, the isotropic behavior is completely independent of the anisotropic one. Then, the claim that one can 
treat the isotropic and anisotropic cases on the same RG grounds \cite{Di2} is not consistent \cite{AL2}.

To summarize, we used a new renormalization group treatment for the $m$-fold Lifshitz critical behavior which 
constitutes a conceptual development in the comprehension of this sort of criticality. By using two 
different external momenta scales for the anisotropic behavior (perpendicular and parallel to the competition axes, 
respectively), we were able to derive two new sets of independent scaling laws, generalizing previous RG treatments. 
To our knowledge, we found for the first time new scaling relations for the isotropic critical behavior for $m=d$ close 
to 8. A detailed description of the contents presented in this letter as well 
as a thorough description of all the critical exponents at least at 
two-loop order for the isotropic and anisotropic behavior will be given in a 
subsequent publication \cite{Lei}. The approach described here might be useful to treat Lifshitz points of generic character, when one allows arbitrary 
even momentum powers (greater than 2) in the Lagrangian (1), as well as other general field theories with 
higher-derivative appearing in other physical contexts \cite{8}.

\bigskip
{\bf Acknowledgments}

The author would like to thank Dr. M. D. Tonasse for a critical reading of the manuscript, and Dr. L. C. de Albuquerque for 
useful discussions. Financial support from FAPESP, grant number 00/06572-6 is gratefully acknowledged.

\end{document}